# VULNERABILITY SCANNERS: A PROACTIVE APPROACH TO ASSESS WEB APPLICATION SECURITY

Sheetal Bairwa[1], Bhawna Mewara[2] and Jyoti Gajrani[3]

[1,2,3]Department of Information Technology, Government Engineering College, Ajmer

## *ABSTRACT*

*With the increasing concern for security in the network, many approaches are laid out that try to protect the network from unauthorised access. New methods have been adopted in order to find the potential discrepancies that may damage the network. Most commonly used approach is the vulnerability assessment. By vulnerability, we mean, the potential flaws in the system that make it prone to the attack. Assessment of these system vulnerabilities provide a means to identify and develop new strategies so as to protect the system from the risk of being damaged. This paper focuses on the usage of various vulnerability scanners and their related methodology to detect the various vulnerabilities available in the web applications or the remote host across the network and tries to identify new mechanisms that can be deployed to secure the network.*

## 1. INTRODUCTION

With the emergence of information technology, the security aspect of the users has become a more concerned factor. Since most of the software developers are not aware of various security measures to be introduced into the system as their motive is just to make the software application run in a desired state without taking into consideration the flaws that the programming language might have introduced into the system; to protect the users from the risk of being attacked by any unauthorised access, it becomes significantly more important to devise new strategies and methodologies that will consider the security breaches to which the user is prone to. Not only the software developed with flaws makes the user vulnerable to attacks, most often network also becomes a key factor by compromising the security aspect of the users.

Assessing and eliminating the vulnerabilities requires the knowledge and deep understanding of these vulnerabilities. It becomes necessary enough to know the basic idea that works behind these vulnerabilities such as what makes them to appear in the system, what flaws need to be corrected to make the system free from these vulnerabilities, what alternatives can be further devised for these vulnerabilities so that in future, their risk can be reduced and many more.
Various methods have been deployed to identify these vulnerabilities and appropriate steps are taken. Strategies such as static analysis, attack graph generation and its analysis, usage of vulnerability scanners are some of them. However, the use of vulnerability scanners to detect the vulnerabilities is quite prominent today. They play a significant role in the generation of attack graphs.

Our work involves study of various port scanners and vulnerability scanners, scanning of various online web applications and remote host using these scanners. We analysed various vulnerabilities and make a comparison of various scanners based on their capability to identify these vulnerabilities.

Section 2 explains various techniques developed before the usage of vulnerability scanners. Section 3 describes various vulnerability scanners in detail with the results, when applied on various websites. Comparative study of various scanners is given in Section 4.

## 2. TECHNIQUES FOR VULNERABILITY SCANNING

### 2.1 STATIC ANALYSIS

Static analysis is a fast and reliable technique. It has been considered as an efficient method in detecting the vulnerabilities [3].This technique focuses on the analysis of program structure using various means. It emphasizes on the analysis of the code of the program in order to detect the flaws present in it.

Some of the techniques included in static analysis are lexical analysis, type inference, constraint analysis and many more. Lexical analysis focuses on the semantics of the program structure; the program structure is divided into modules and then each module is compared with the loophole library in order to detect any flaws present in the system. Type inference is related to the data type rules for the variable. It determines whether the variables used in the program are in sync with the type to which they relate. Constraint analysis is a two-step process. It involves- constraint generation and constraint solution [1].

Many tools based on the techniques mentioned above are developed. The first tool developed was FlexeLint. It uses pattern matching algorithm to detect flaws. Other tools developed are ITS4, SPLINT, UNO, FindBugs, Checkstyle, ESC/Java, and PMD. ITS4, Checkstyle and PMD are based on lexical analysis; SPLINT is based on rule checking; UNO is based on model checking; ESC/Java is based on theorem proving and FindBugs is based on both lexical and dataflow analysis [1].

These tools have been evaluated by analysing their performance in terms of false positives and false negatives. Many of them have low false positives, some produce accurate results and many witnessed high false negatives. Hence, static analysis techniques have many demerits associated with them. For instance, a loophole library or database is maintained which is used to validate the vulnerabilities found in the program; however if an unknown vulnerability is detected, then it is not possible to compare it with the predefined loophole library for its validation [1].

Thus, to resolve the deficiencies associated with the static analysis, an approach was suggested that involved combining the dynamic detection strategy with static analysis.

### 2.2 ATTACKGRAPH ANALYSIS

Attack graph is defined as the succinct representation of all the paths followed by an attacker in a network to achieve its desired state. The desired state may involve damaging the network, stealing the network packets or gaining a complete access over it to determine what is going in the network.

International Journal on Computational Sciences & Applications (IJCSA) Vol.4, No.1, February 2014

Network security is a key aspect of security concern and many ways have been identified to protect it. The recent approach that has been included is the use of attack graphs. Attack graph has become the most widely used approach with reference to network security.

Attack graphs help to determine the security weaknesses that lie in the network. System administrators use it to analyze the network for its weaknesses that may allow an attacker to exploit it and gain control over the network [2]. Attack graphs are usually large enough as they represent the complete network with its underlying weaknesses, hence they are quite complex to understand and analyse. Both the generation and analysis of attack graph are significant for protecting the network from security breaches.

The most common approach to generate an attack graph requires the analysis of vulnerabilities that lie in the network and then using an attack graph generator, attack graphs can be generated [4]. The vulnerabilities could be identified with the help of various vulnerability scanners that are designed for this purpose only. Specifically, Nessus is extensively used for the identification of the underlying vulnerabilities.

Various other techniques have already been proposed for generating an attack graph as well as for their analysis. For instance, adjacency matrix clustering algorithm makes the complex attack graph simpler enough. It combines the blocks having similar attack graph pattern. The matrix represents the attack reachability within one step. For multiple steps, matrix is raised to a higher power level [13].

Ranking algorithm is another approach, based on the rank of the attack graphs. The rank decides the priority of an attack graph that is more applicable to attacker [14]. Another approach is a game theoretic approach where the attacker and network administrator are considered as two players and a Nash equilibrium is applied that gives the administrator an idea of attacker's strategy and helps him to plan to do something in order to protect the network [12].

Table 1 above compares the various attack graph generation and analysis techniques and illustrates the advantages and disadvantages of each [2].

| Technique | Author | Merits | Demerits |
|---|---|---|---|
| Clustered adjacency matrix | Steven Noel Sushi Jajodia | Automatic, parameter-free, and scales linearly with problem size | Need to calculate highest level of adjacency matrix for multistep reachability |
| Hierarchical aggregation | Steven Noel Sushi Jajodia | Framework useful for both computational and cognitive scalability | The process of interactive de-aggregation is potentially tedious to determine low level details |
| Minimization analysis | S. Jha O. Sheyner J. Wing | Identifies the smallest set of countermeasures required to prevent all possible attack paths | Approach is limited to Directed Acyclic Graph |
| Ranking graph | Vaibhav Mehta C. Bartzis Haifeng Zhu Edmund Clarke J. Wing | Ease and flexibility of modelling | Difficult for security manager to make decision on actions to protect network |
| Game theoretic | K.W. Lye Jeannette Wing | Allows to know more about attacker's attack strategies | Full state space is extremely large. |

Table 1: Comparison of the attack graph techniques

115

## 3. VULNERABILITY SCANNERS

A large number of applications are becoming online, but how secure are these products is a matter of concern as it is related to the user's security who will be ultimately using the application. Thus, it becomes necessary to find out vulnerabilities present in the software application that may cause a severe risk to the user's security [5].

Vulnerability assessment means identifying the vulnerabilities in the system before they could be used by anyone else with bad intentions of harming the network. This is a proactive approach where the vulnerability is found and is dealt with accordingly before anyone comes to know about it. More emphasis has always been laid on the firewall protection but the internal functionality does matter. Vulnerability assessment is not only performed on a particular application but it even correlates the platform on which the application is being run, middleware, operating system being used etc. It takes into consideration all the factors that can provide the correct answer for the assessment of the vulnerability and security of the system. Therefore, vulnerability scanners are used to scan the network system and/or the software applications.

Scanning can be of two types:

a) **Passive Scanning**: In passive scanning, it is determined whether a tool can enlist the vulnerabilities by considering the existing network.
b) **Active Scanning**: In active scanning, it is determined whether the queries can be made to the network for the vulnerability.

Different categories of scanner are:

a) **Port Scanners**: Port scanners are used to scan the ports for determining the open and closed ports, operating system, services offered.
b) **Application Scanners**: Application scanners are used to assess a specific application on the network in order to track its weaknesses that can be further used to cause the risk to the system.
c) **Vulnerability Scanners**: Vulnerability scanners are the ones that find out the vulnerabilities in the system which if accessed by a malicious user or hacker can put the whole network system at risk.

Penetration testing is the other concept that follows the vulnerability assessment. With penetration testing, it is possible to make use of the loopholes or vulnerabilities to gain an unauthorised access. It validates how effectively the system can respond to the real life attacks.

OWASP (Open Web Application security Project) focuses on providing the better security of the software. It has enlisted commonly critical vulnerabilities that the application may be prone to. These vulnerabilities when exploited provide the risk of losing security and confidentiality. For instance, Injection vulnerability occurs due to the execution of a command or query for an untrusted data; Broken Authentication and Session Management, due to improper implementation of an application risks the user's confidentiality. Cross Site Scripting, commonly referred as XSS is another flaw in which attacker injects malicious script into web pages viewed by users and also to bypass access controls. Insecure Direct Object References, in which developers unknowingly leave some holes which give a chance to attackers to access and manipulate directory, database key. Cross Site Request Forgery or CSRF, is an attack where user is forged to click on a link that is intuitively designed to steal the cookies and other private details of the user. Sensitive data exposure is another area of vulnerability where the sensitive data such as credit card details, authentication credentials etc. are not secured which helps an attacker to conduct the fraud [15].

International Journal on Computational Sciences & Applications (IJCSA) Vol.4, No.1, February 2014

Next subsections discuss various scanners and the results obtained by scanning various web applications using these scanners.

## 3.1 NMAP

Nmap is a port scanner that is used to scan the ports. It takes an IP address or the host name and then finds the basic information related to it. If an IP address is provided, it then finds the host to which it belongs to. It also finds the number of ports that are running on that particular host, number of ports that are opened, number of closed ports, services provided by those ports, for instance, whether services are TCP-oriented or FTP-oriented [10]. It even predicts the type of operating system being used on that particular host. The topology of the scanned host is recorded in the graphical format which shows the various gateways through which the local machine accesses that particular remote host.

Considering the ports that are opened, an attack can be designed in order to have an unauthorised and a legitimate access to the host with a goal set in mind. Moreover, if the opened ports are providing the services which are TCP-oriented or FTP-oriented, it becomes easy to gain access to the host.

A number of various sites have been scanned using NMAP. The figure below depicts the results obtained after scanning RTU website.

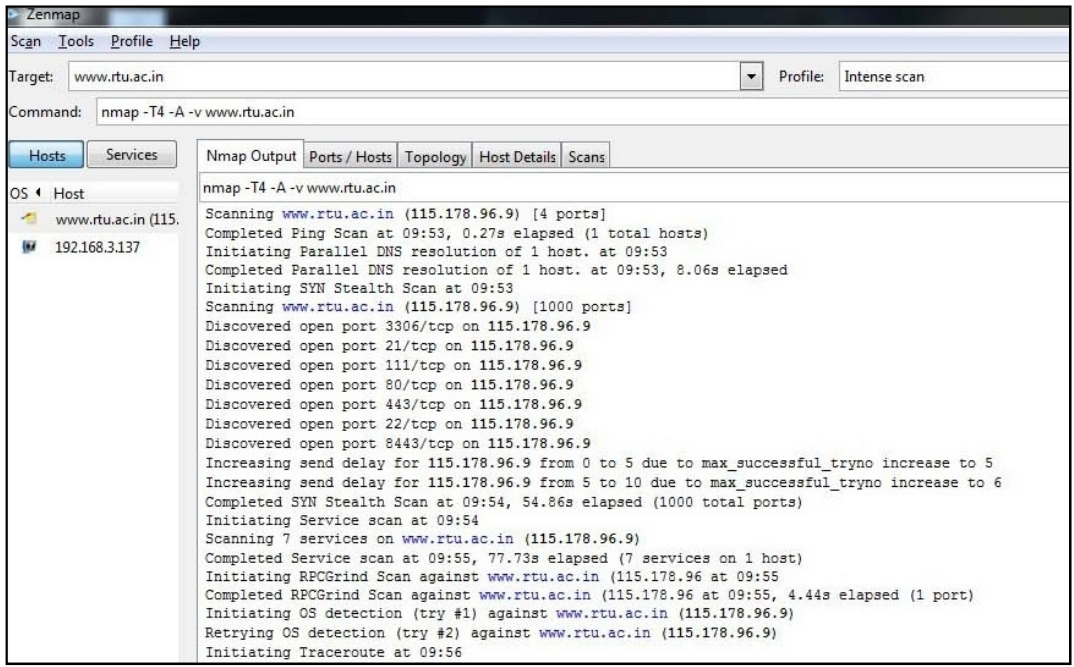

Figure1. Nmap basic output for RTU website

Figure 1 shows the basic details of RTU website including the IP address, number of total ports available, number of open ports discovered, performing RPCGrind scan and much more other relevant details.

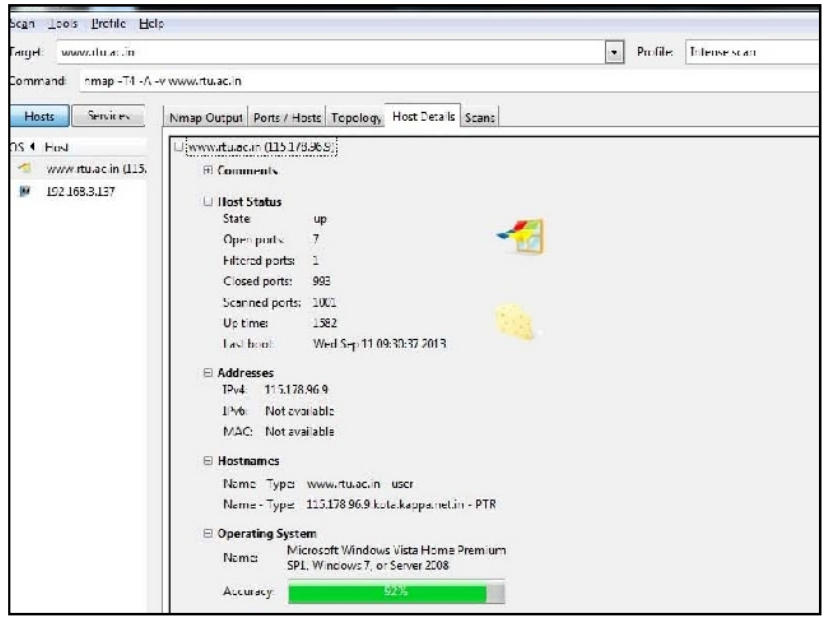

Figure2. Host details of RTU

Figure 2 outlays the host details of RTU website which includes the host status that depicts the number of total ports scanned, number of open ports available, number of filtered ports. It shows IPv4 address of the website ;IPv6 and MAC address are not available for this website. Further, the type of operating system used and its accuracy of being correct is also illustrated. In this case, types of operating system detected may be Microsoft Windows Vista Home Premium SPI, Windows 7or Server2008. The accuracy with which this result has been obtained is 92% approximately.

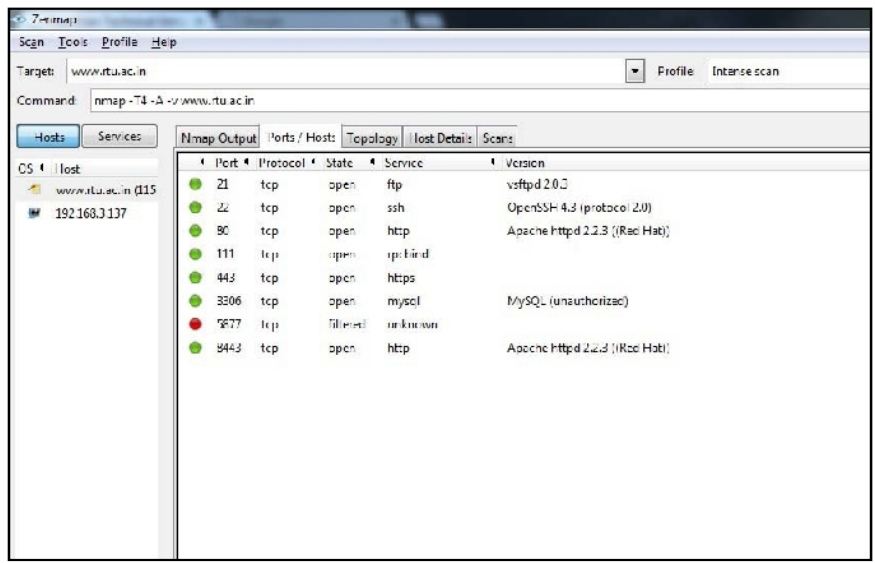

Figure3. Details of closed and open ports of RTU

International Journal on Computational Sciences & Applications (IJCSA) Vol.4, No.1, February 2014

Figure 3 shows the list of open ports on RTU website. It depicts the port number, protocol used on that port, its state of being open or closed or filtered, type of service provided on that port and the version details. For instance, port 21 is in open state where TCP protocol is used and the service provided is FTP.

## 3.2 NESSUS

Nessus is a vulnerability scanner that lists the various vulnerabilities present in the remote host. It provides both internal and external scan. Internal scan is related to hosts within a particular router. External scan involves the hosts outside the particular router (a remote host). Web application test is also performed in the scanner. Either the scanning can be done at the first instance provided or a template can be created first for a particular host and then it can be launched to run the scan against that host. Multiple scanning of the hosts can be done at once. The vulnerability found by Nessus exists in four different types of severity- High, Medium, Low and Informal [7].

Results are also saved as soon as the scan of a particular host is completed. The results are provided in two different ways- vulnerabilities by plug-ins and vulnerabilities by host. The former category first classifies all the vulnerabilities found during scan, and then shows the hosts affected by those vulnerabilities. Using the generated report, problems can be identified and fixed easily. The latter category identifies all hosts found during the scan and their associated vulnerabilities. This report addresses distinct issues associated with reliable hosts, PCI scans, follow-up scans, and targeted assessments. PVS real-time scanning completes Nessus active scanning by providing continuous network evaluation and bridges security gaps between scans. The results can be exported in any desired format (e.g. PDF, HTML, CSS etc).

Nessus is based on client –server architecture. Each session is controlled by the client and the test is run on the server side.

More than 100 websites have been scanned using Nessus. The figure below depicts the results obtained for Hebron website.

Figure4. Vulnerability details for Hebron website using Nessus

International Journal on Computational Sciences & Applications (IJCSA) Vol.4, No.1, February 2014

Figure 4 shows the scan results for Hebron website using Nessus scanner. It shows total 29 vulnerabilities for this particular website- 0 Critical, 1 High, 3 Medium, 2 Low and 23 Informal. The range of either being high, medium, low or informal type is also given. For instance, FTP privileged port bounce scan is belongs to high category ranging to 7.5 with its plugin ID given as 10081.

The report generated provides the description for all the vulnerabilities that occurred in the scanning process with its appropriate solution. For example, the report of Hebron website states that FTP privileged port bounce scan belonging to plugin ID 10081means that the remote FTP server is possibly vulnerable to FTP bounce attack i.e. forcing the remote FTP server to connect to third parties providing the intruder an opportunity to make use of their resources and compromise the security of the website.

## 3.3 ACUNETIX WVS

Acunetix WVS is an exploit analysis tool for performing web security audits. The criteria on which Acunetix WVS work includes- target specification, site crawling and structure mapping and pattern analysis.

   a) **Target Identification**: WVS checks target(s) with active web server, and therefore, host any web application. Information is collected regarding web-technologies used, web server-type and responsiveness for appropriate filtering tests.

   b) **Site Crawling and Structure Mapping**: The index file of web application is fetched first, determined by the URL (e.g., http://192.168.1.128:80/ will load the main index.html). Received responses are parsed to get links, forms, parameters, input fields, and client side scripts that builds a list of directories and files inside the web application.

   c) **Pattern Analysis** is executed against the web application.

Various web applications have been scanned using Acunetix WVS. The figure shown below depicts the result obtained after scanning Air India Website.

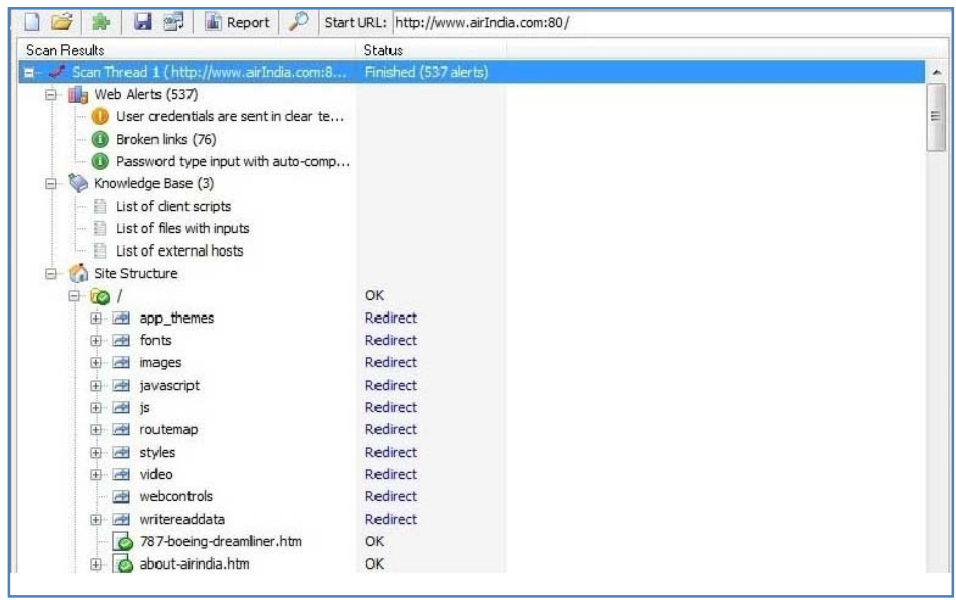

Figure5. Crawling details for Air India website using Acunetix

Figure 5 shows the crawling structure of Air India website obtained after scanning it using Acunetix WVS. Crawling, in general, refers to navigate all the pages of a complete web application. It enlists all the various portions of websites that have been scanned and identifies the vulnerability which may be present in any of those crawled pages.

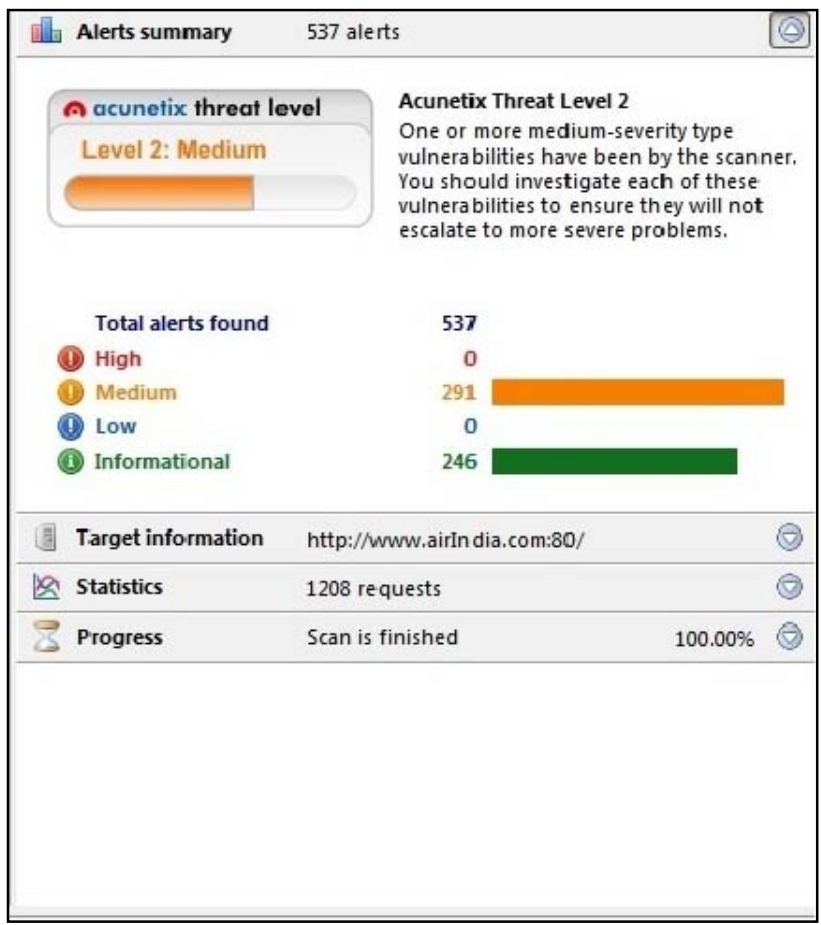

Figure6. Vulnerability Alert Summary Details of Air India website using Acunetix WVS

Figure 6 shows the summary of vulnerabilities found during the scan of Air India website. It provides the target information, types of vulnerabilities detected and the vulnerability threat level. In this case, total 537 alerts are found- 291 Medium and 246 Informal. As such, no high category vulnerability is detected and the preferred Acunetix threat level is Level 2: Medium.

### 3.4 NIKTO

Nikto is a command based tool that is also used to scan the specific targets. It requires having the Perl language installed in the system since the functionality is based on this language. It performs the security check against dangerous files/CGI problems on servers. Attackers look for web server vulnerabilities to gain access of everything from insecure WordPress implementation to outdated Apache servers.

Nikto is free and open source Web server security scanner therefore IT security teams can better understand the server security at their enterprises and take positive steps toward shielding and

International Journal on Computational Sciences & Applications (IJCSA) Vol.4, No.1, February 2014

upgrading systems. The tool is able to find the scamp servers that weren't set up by the enterprise and reveals vulnerabilities. It can also perform the security check against over 65,000 dangerous files/CGI and 1,250 outdated servers [11].

### 3.5 BURPSUITE

Burp is a proxy based tool package. It consists of various functional specifications. To start working with Burp, it first requires setting the proxy in the browser whichever is being used as 127.0.0.1. After the proxy is set in the browser, Burp is ready to begin with.

Burp window involves many tab specifications such as Proxy, Intruder, Spider, Repeater, Sequencer and Scanner etc. where each tab has its own sub tabs. For instance, Proxy tab has three sub tabs-Intercept, Proxy, Options.

Proxy tab is used to set the proxy and configure it. The Intercept sub tab within it remains on at this time. A Xampp server is installed in the system which provides the Mutillidae server that is developed with the idea of testing the applications. Through this, you can identify the username and password for a particular user provided that Intercept tab is off at that time when you are trying to access it from Mutillidae.

Intruder tab is used to automate customized attacks against web applications to detect and exploit all common vulnerabilities. Spider tab provides the crawling feature in the web application test. Repeater tab is used to modify HTTP requests manually and analyses their responses [8].

Scanner performs the scanning of the hosts. With trial version, The Scanner feature is not available. A full professional version needs to be purchased in order to perform the scanning. Scanning involves testing the hosts for the vulnerabilities present in it. It identifies the type of vulnerability and its severity.

## 4. COMPARISON OF SCANNERS

Table 2 shows comparative view of the tools mentioned above on the basis of the vulnerabilities these tools detect.

Table2. Comparative view of the vulnerabilities detected by the scanners

| Vulnerabilities | Nmap | Nessus | Acunetix WVS | Nikto | BurpSuite |
|---|---|---|---|---|---|
| SQL Injection | | | | | |
| Improper Error Management | | | | | |
| Cross site Scripting | | | | | |
| Rogue Servers | | | | | |
| Denial of Service | | | | | |
| Remote Code Execution | | | | | |
| Format String Identifier | | | | | |
| IIS.printer | | | | | |
| DCOM | | | | | |

International Journal on Computational Sciences & Applications (IJCSA) Vol.4, No.1, February 2014

The table compares the different scanners for the different vulnerabilities. As seen from the table, Nessus is the only scanner that has detected most of the listed vulnerabilities followed by Acunetix WVS and Burp.

Various web applications are scanned using Nessus, Retina, Netrecon and ISS in [5] and comparison is made between them as shown in table 3.

Table3. Comparison of different scanners given by [5]

| Vulnerabilities | Retina | Nessus | Netrecon | ISS |
|---|---|---|---|---|
| RPCBind | | | | |
| Finger &SSH | | | | |
| LSSAS.exe | | | | |
| SQL Injection/Preauthentication | | | | |
| XSS | | | | |
| WWW, using cmd.exe | | | | |
| SSH on high port number | | | | |
| IIS.printer | | | | |
| DCOM | | | | |

Comparing Table 2 and Table 3, it is observed that Nessus is a commonly used tool. Our paper in contrast to [5], focuses on more recent tools and vulnerabilities, as the study in [5] is done in year 2006. In table 2, Nessus has come out with the best scanning capabilities while in table 3, Retina has proved itself to be good among all others. This is due to the choice of vulnerabilities selected for comparison.

## 5. CONCLUSION

Various techniques can be used to list the vulnerabilities present in the web applications or remote host. Vulnerability assessment plays a significant role in securing the network system. Our observations show that different scanners detect different type of vulnerabilities but a single tool is not capable of detecting all type of vulnerabilities. This paper addressed various tools used for scanning vulnerabilities and their comparative study. We identified what vulnerabilities a specific tool is capable of detecting by running each on a number of web applications

BurpSuite has many features incorporated within it which is not available in other tools and hence can be integrated with the other tool that works differently and produces different results. Sometimes a tool detects the vulnerabilities that may not be detected by the other tools, so it will prove bonus if various tools integrate with each other as the number of vulnerabilities detected will be the total sum of all the vulnerabilities detected by each tool which will be greater than the number found by each tool individually. Although the task is tedious but with thorough understanding of the different criteria on which the tools work may lessen the burden of complicacy of implementing the same. Our research focuses on the implementation of a test-bed where different scanners will be combined on the basis of their capabilities.

123

**Authors**

**Sheetal Bairwa**completed her B.Tech degree in from Information Technology from Government Women Engineering College under Rajasthan Technical University, Kota, India in 2012.She is pursuing the M. Tech in Information Technology from Government Engineering College, Ajmer under Rajasthan Technical University, Kota, India.

Her research interests include Network security and Information security.

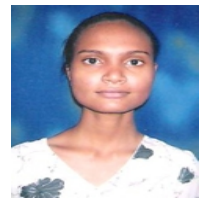

**Bhawna Mewara**completed her B.Tech degree in Computer Engineering from Arya Institute of Engineering and Technology under Rajasthan Technical University Kota,India in 2012.She is pursuing the M. Tech in Information Technology from Government Engineering College, Ajmer under Rajasthan Technical University, Kota, India.

Her research interests include Network security and Information security.

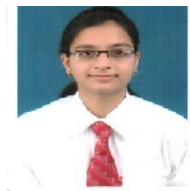

**Jyoti Gajrani** graduated from ModyCollege of Engineering and Technology, Lakshmangarh, under Rajasthan University, Jaipur, Rajasthan, India in 2004. She received her M.Tech in Computer Engineering from IIT Bombay, India in 2013.

Her research interests include Information Security, Network Security, Databases, Distributed Applications and Computer Architecture.

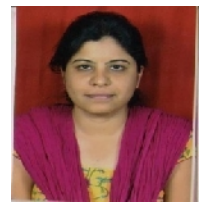